\documentstyle[pra,aps]{revtex}
\begin{document}
\draft
\title{Brownian Motion: the Quantum Perspective}
\author{Bassano~Vacchini}
\address{Dipartimento di Fisica  
dell'Universit\`a di Milano and Istituto Nazionale di Fisica  
Nucleare, Sezione di Milano, Via Celoria 16, I-20133, Milan,  
Italy
\\
Reprint requests to B.~Vacchini; E-mail: bassano.vacchini@mi.infn.it
}  
\date{\today}
\maketitle
\begin{abstract}
We briefly go through the problem of the quantum description of
Brownian motion, concentrating on recent results about the connection
between dynamics of the particle and dynamic
structure factor of the medium.
\end{abstract}
%\pacs{05.40.Jc, 05.30.Fk, 05.30.Jp, 03.65.Yz}
\pacs{\textit{Keywords}: Quantum Brownian Motion, Dynamic Structure
  Factor, Fokker-Planck, Lindblad}
%%%%%%%%%%%%%%%%%%%%%%%%%%%%%%%%%%%%%%%%%%%%%%%%%%%%%%%%%%%%%
%%%%%%%%%%%%%%%%%%%%%%%%%%%%%%%%%%%%%%%%%%%%%%%%%%%%%%%%%%%%%
Early in the nineteenth century, the biologist Robert Brown wrote a
paper on the observation of the random motion of a large particle
immersed in a fluid~\cite{Brown}, a paper which received wide
attention, leading to the phenomenon being named after him. Despite
the raised interest, the first correct theoretical description of the
phenomenon, which had previously been tentatively explained in terms
of irregular heating due to incident light or some kind of electrical
forces, was only given after almost a century on the basis of the
so-called \textit{random walk problem} in a number of papers
(beginning 1905~\cite{Einstein}) by Albert 
Einstein, who was looking for a way to confirm the atomic nature of
matter. These studies led to a relation between the macroscopic
diffusion coefficient and the atomic properties of matter known as
Einstein's relation, which links the irreversible nature of the
phenomenon to the mechanism of molecular fluctuations, thus providing
the first example of fluctuation-dissipation relation, the key point
being the connection between the dynamics of the Brownian particle and 
statistical mechanics properties of the fluid. 

The description of the phenomenon at
classical level relies on differential equations of the Fokker-Planck
type for the different distribution functions\cite{Risken}, and in
particular the velocity distribution of the Brownian particle obeys an 
equation of the form
\begin{equation}
\label{fokker}
  \frac{\partial p({v},t)}{\partial t}
=
\eta 
\frac{\partial}{\partial v}[v p({v},t)]
+
D_v \frac{\partial^2}{\partial v^2} p({v},t)
,
\end{equation}
where $D_v = \eta / M\beta$ and $\eta$ is the friction constant. It is
however not immediately obvious how to tackle the problem in the
quantum case, since due to dissipation and irreversibility one is not
dealing with a Hamiltonian system and therefore the standard
quantization procedure fails. At this point one can either resort to
new quantization schemes, which might lead to meaningful results for
this particular type of problems\cite{Dekker}, or more consistently
consider a system reservoir approach inside a thorough quantum
mechanical formalism\cite{Spohn}, aiming at the description of the
subdynamics of the Brownian particle with respect to the fluid. A
first naive guess of the possible result may be obtained applying the
correspondence principle to (\ref{fokker}), written however in terms
of momenta, substituting to the classical variables the operators
${\hat {{\sf x}}}$ and ${\hat {{\sf p}}}$ acting in the usual way in
the Hilbert space of the particle, thus obtaining in the Schr\"odinger
picture a structure of the form
\begin{equation}
  \label{guess}
         {  
        d {\hat \varrho}  
        \over  
                dt  
        }  
        =
        -
        {i\over\hbar}
         [  
        {\hat {{\sf H}}} ,
        {\hat \varrho}
        ]           
        -
        {i\over\hbar}
        \gamma
        \left[  
        {\hat {{\sf x}}} ,
        \left \{  
        {\hat {\mbox{\sf p}}},{\hat \varrho}
        \right \}  
        \right]     
        -
        {
        D_{pp}  
        \over
         \hbar^2
        }
        \left[  
        {\hat {{\sf x}}},
        \left[  
        {\hat {{\sf x}}},{\hat \varrho}
        \right]  
        \right]  
, 
\end{equation}
where for the sake of simplicity we have here restricted ourselves to 
a one dimensional notation.
This somehow expected result was in fact obtained in a pioneering work 
by Caldeira and Leggett~\cite{Caldeira}, using a Feynman path
integral formalism for the description of a particle coupled to a bath
of harmonic oscillators. In particular their result takes the form
\begin{equation}
  \label{caldeira}
          {  
        d {\hat \varrho}  
        \over  
                dt  
        }  
        =
        -
        {i\over\hbar}
         [  
        {\hat {{\sf H}}} ,
        {\hat \varrho}
        ]           
        -
        {i\over\hbar}
        \gamma
        \left[  
        {\hat {{\sf x}}} ,
        \left \{  
        {\hat {\mbox{\sf p}}},{\hat \varrho}
        \right \}  
        \right]     
        -
        {
        1  
        \over
        \hbar^2
        }
        \frac{2M\gamma}{\beta}
        \left[  
        {\hat {{\sf x}}},
        \left[  
        {\hat {{\sf x}}},{\hat \varrho}
        \right]  
        \right]  ,
\end{equation}
where $\beta=1/kT$ is the inverse of the temperature and $\gamma$ is a
phenomenological parameter linked to the bath properties. 
It was however later realized 
that this master equation does not always preserve positivity of the
statistical operator~\cite{Ambegaokar,Pechukas2}: in fact it does not have a
completely positive structure~\cite{LindbladQBM,AlbertoQBM}, 
complete positivity being for this class of systems
equivalent to positivity~\cite{Talkner}. A completely positive time
evolution is in fact driven by a master equation having a Lindblad
structure~\cite{Lindblad}
\begin{equation}
  \label{lindblad}
            {  
        d {\hat \varrho}  
        \over  
                dt  
        }  
        =
        -
        {i\over\hbar}
        [
        {{\hat {\mbox{\sf H}}}}
        ,{\hat \varrho}
        ]
        -
        {
        1
        \over
         2\hbar
        }
        \sum_i
        \{\hat{V}_i^\dagger \hat{V}_i,{\hat \varrho} \}
        +
        {
        1
        \over
        \hbar
        }
        \sum_i
        \hat{V}_i{\hat \varrho}\hat{V}_i^\dagger
,
\end{equation}
which in the case of Brownian motion, in order to give a friction
force proportional to velocity, takes the following form, with
contributions at most bilinear in the operators ${\hat {{\sf x}}}$ and
${\hat {{\sf p}}}$, according to a choice of generators of the form
$\hat{V} = a{\hat {{\sf x}}}+b{\hat {{\sf p}}}$~\cite{Sandulescu}
\begin{eqnarray}
\label{sandulescu}
        {  
        d {\hat \varrho}  
        \over  
                dt  
        }  
        =
        &-&
        {i\over\hbar}
        [
        {\hat {\mbox{\sf H}}}
        ,{\hat \varrho}
        ]
        -
        {i\over\hbar}\mu
        [
        {\hat \varrho},
                \left \{  
        {\hat {\mbox{\sf x}}},{\hat {\mbox{\sf p}}}
        \right \}  
        ]
        -
        {i\over\hbar}
        \gamma
        \left[  
        {\hat {{\sf x}}} ,
        \left \{  
        {\hat {\mbox{\sf p}}},{\hat \varrho}
        \right \}  
        \right]   
        \\
        &-&
        {
        D_{xx}
        \over
         \hbar^2
        }
        \left[  
        {\hat {\mbox{\sf p}}},
        \left[  
        {\hat {\mbox{\sf p}}},{\hat \varrho}
        \right]  
        \right]  
        -
        {
        D_{pp}  
        \over
         \hbar^2
        }
        \left[  
        {\hat {{\sf x}}},
        \left[  
        {\hat {{\sf x}}},{\hat \varrho}
        \right]  
        \right]  
        +
        2{
        D_{xp}  
        \over
         \hbar^2
        }
        \left[  
        {\hat {{\sf p}}},
        \left[  
        {\hat {{\sf x}}},{\hat \varrho}
        \right]  
        \right]  
        \nonumber
        ,
\end{eqnarray}
the coefficients being constrained by the following requirements:
\begin{eqnarray}
  \label{vincoli}
 &&   D_{pp}>0, \quad  D_{xx}>0, 
\nonumber 
   \\
  &&   D_{xx}D_{pp}-D_{xp}D_{px}\geq
  (\gamma\hbar/2)^2  
,
\end{eqnarray}
which ensure complete positivity. Equation (\ref{caldeira}) is not of
Lindblad form, since $D_{xx}=0$, and quite a lot of work has been done in
order to cope with this difficulty~\cite{Tannor-97}, both at
fundamental and phenomenological level, usually considering an
harmonically bound particle interacting with a bath of harmonic
oscillators, leading to a typical correction of the form
$$D_{xx}=\chi \frac{\beta\hbar^2}{M}\gamma,$$ with
$\chi\geq{1}/{8}$, corresponding to a higher order correction in 
the high temperature limit actually considered by Caldeira and
Leggett.

Most recently a microphysical scattering theory derivation of a master 
equation driving the subdynamics of a test particle interacting
through collisions with a fluid has been considered and applied to the 
specific case of the quantum description of Brownian
motion~\cite{art1,art3,art4}. Let us note that the particular case of
Brownian motion bears some peculiar features with respect to the more
general dissipative systems described by~\cite{Sandulescu}, as
stressed in~\cite{Tannor-97,reply} and often neglected in the
literature. The main result of~\cite{art4} is the following structure
of master equation for the subdynamics of the test particle immersed
in a homogeneous fluid of not necessarily free particles:
\begin{eqnarray}
  \label{core}
        {  
        d {\hat \varrho}  
        \over  
                      dt
        }  
        =
        &-&
        {i \over \hbar}
        [{\hat {{\sf H}}}_0
        ,
        {\hat \varrho}
        ]
        \\
        &+&
        {2\pi \over\hbar}
        (2\pi\hbar)^3
        n
        \int d^3\!
        {\bbox{q}}
        \,  
        {
        | \tilde{t} (q) |^2
        }
        \Biggl[
        L({\bbox{q}},{\hat {{\sf p}}},{\hat
        {{\sf x}}})
        {\hat \varrho}
        L^{\scriptscriptstyle\dagger}
        ({\bbox{q}},{\hat {{\sf p}}},{\hat {{\sf x}}})
        -
        \frac 12
        \left \{
        L^{\scriptscriptstyle\dagger}
        ({\bbox{q}},{\hat {{\sf p}}},{\hat {{\sf x}}})
        L({\bbox{q}},{\hat {{\sf p}}},{\hat
        {{\sf x}}}),
        {\hat \varrho}
        \right \}
        \Biggr],
        \nonumber
\end{eqnarray}
where ${\hat {{\sf H}}}_0$ is the Hamiltonian of the free particle,
${\hat \varrho}$ its statistical operator, $n$ the particle density in 
the fluid, $\tilde{t} (q)$ the Fourier transform of the T
matrix describing two-particle collisions evaluated for the
transferred momentum $q$ and $$L({\bbox{q}},{\hat {{\sf p}}},{\hat
        {{\sf x}}})
        =
        e^{{i\over\hbar}{\bbox{q}}\cdot{\hat {{\sf x}}}}
        \sqrt{
        S({\bbox{q}},{\hat {{\sf p}}})
        },$$
with $S({\bbox{q}},{\hat {{\sf p}}})$ the dynamic structure function
of the medium. Expression (\ref{core}), with its typical Lindblad
structure, can be obtained under a suitable interplay between energy
dependence of the dynamic structure 
factor and quasi-diagonality of the matrix elements of
${\hat \varrho}$ in the momentum representation, valid if one works on 
a time scale $\tau$ over which the subdynamics of the particle is
suitably slow. The dynamic structure factor, 
usually given in terms of energy
and momentum transfer ($E$ being the energy transfer corresponding to
scattering from state 
${\bbox{p}}$ to state ${\bbox{p}}' =
{\bbox{p}}+{\bbox{q}}$), is the Fourier transform of the two-point
density correlation function of the medium~\cite{Lovesey}
\begin{equation}
  \label{dsf}
  {S} ({\bbox{q}},E)=
        {  
        1  
        \over  
         2\pi\hbar
        }  
        {
        1
        \over
         N
        }
        \int dt 
        {\int d^3 \! {\bbox{x}} \,}        
        e^{
        {
        i
        \over
         \hbar
        }
        (E t -
        {\bbox{q}}\cdot{\bbox{x}})
        }  
        {\int d^3 \! {\bbox{y}} \,}
        \left \langle  
         N({\bbox{y}})  
         N({\bbox{x}}+{\bbox{y}},t)
         \right \rangle
\end{equation}
and can be measured in terms of scattering experiments, being strictly 
connected to the inelastic scattering cross-section of a microscopic
probe off a macroscopic sample~\cite{vanHove}. Expression (\ref{dsf})
depends on the statistical mechanics properties of the system and in
particular on its equilibrium fluctuations, so that according to
(\ref{core}) the dynamics of the particle depends on the density
fluctuations of the medium, thus putting into major evidence through
the appearance of the quantum correlation function (\ref{dsf}) one of the main 
ideas of the classical approach of Einstein to the problem. To proceed further
we have to give a specific example of dynamic structure factor 
and we consider the case of
a free gas. To restrict to Brownian motion one takes a particle of mass
$M$ much heavier than the mass $m$ of the particles making up the
gas. In the case of Boltzmann particles this leads to~\cite{art3}
        \begin{eqnarray}
        \label{qbm}
        {  
        d {\hat \varrho}  
        \over  
                dt  
        }  
        =
        &-&
        {i\over\hbar}
        [
        {{\hat {\mbox{\sf H}}}_0}
        ,{\hat \varrho}
        ]
        \\
        &+&
        z
        {4\pi^2 m^2 \over\beta\hbar}
        \int d^3\!
        {\bbox{q}}
        \,  
        {
        | \tilde{t} (q) |^2
        \over
        q
        }
        e^{-
        {
        \beta
        \over
             8m
        }
        {{{q}}^2}
        }
        \Biggl[
        e^{{i\over\hbar}{\bbox{q}}\cdot{\hat {{\sf x}}}}
        e^{-{\beta\over 4M}{\bbox{q}}\cdot{\hat {{\sf p}}}}
        {\hat \varrho}
        e^{-{\beta\over 4M}{\bbox{q}}\cdot{\hat {{\sf p}}}}
        e^{-{i\over\hbar}{\bbox{q}}\cdot{\hat {{\sf x}}}}
        - {1\over 2}
        \left \{
        e^{-{\beta\over 2M}{\bbox{q}}\cdot{\hat {{\sf p}}}}
        ,
        {\hat \varrho}
        \right \}
        \Biggr]
        ,
        \nonumber
        \end{eqnarray}
where $z$ is the fugacity of the gas, 
and in the limit of small momentum transfer 
or long wavelength fluctuations, in order to recover the analogous of
the Fokker-Planck equation, with a friction force proportional to
velocity, one comes to
\begin{equation}
  \label{z}
         {  
        d {\hat \varrho}  
        \over  
                dt  
        }  
%        &=&
        =
        -
        {i\over\hbar}
        [
        {{\hat {\mbox{\sf H}}}_0}
        ,{\hat \varrho}
        ]
        -
        z
        \sum_{i=1}^3
        \left \{
        {
        {
        D_{pp}
        \over
         \hbar^2
        }
        \left[  
        {\hat {{\sf x}}}_i,
        \left[  
        {\hat {{\sf x}}}_i,{\hat \varrho}
        \right]  
        \right]
%        \right.
%        \nonumber 
%        \\
%        &&
%        \hphantom{\quad\quad}
        +
%        \left.
        {
        D_{xx}
        \over
         \hbar^2
        }
        \left[  
        {\hat {\mbox{\sf p}}}_i,
        \left[  
        {\hat {\mbox{\sf p}}}_i,{\hat \varrho}
        \right]  
        \right]  
        +{i\over\hbar}
        \gamma
        \left[  
        {\hat {{\sf x}}}_i ,
        \left \{  
        {\hat {\mbox{\sf p}}}_i,{\hat \varrho}
        \right \}  
        \right]
        }
        \right \} 
,
\end{equation}
where the coefficients are fixed at microphysical level by the expressions
        \begin{eqnarray}
        \label{coeff}  
        D_{pp}
        &=&
        \frac 23
        {\pi^2 m^2 \over\beta\hbar}
        \int d^3\!
        {\bbox{q}}
        \,  
        {
        | \tilde{t} (q) |^2
        }
        q
        e^{-
        {
        \beta
        \over
             8m
        }
        {{{q}}^2}
        },
        \nonumber
        \\
        D_{xx} 
        &=& 
        (
        {
        \beta\hbar
        /
            4M
        })^2 D_{pp},
        \quad
        \gamma =
        ({
        \beta
        /
             2M
        })
        D_{pp}.
        \end{eqnarray}
Equation (\ref{z}) corresponds to a Lindblad structure with
$\chi={1}/{8}$, i.e., the minimal modification to
(\ref{caldeira}) which copes with complete positivity. In particular
this implies that (\ref{z}) can be written in terms of a single
generator for each Cartesian direction, as one can see introducing the 
operators $
{\hat {\mbox{\sf a}}}_i=
{
\sqrt{2}
\over
 \lambda_M
}
\left(
{\hat {\mbox{\sf x}}}_i
+{i\over\hbar}
{
\lambda_M^2
\over
                    4
}
{\hat {\mbox{\sf p}}}_i
\right)
$, with
$
\lambda_{M}=\sqrt{\hbar^2 / MkT}
$
the thermal wavelength of the Brownian particle (resulting as usual $
[
{\hat {\mbox{\sf a}}}_i  ,
{\hat {\mbox{\sf a}}}_j^{\scriptscriptstyle\dagger}
]
=\delta_{ij}$), thus coming to
        \begin{eqnarray}
        \label{a}
        {  
        d {\hat \varrho}  
        \over  
        dt
        }  
        =  
        &-&
        {i\over\hbar}
        [
        {{\hat {\mbox{\sf H}}}_0}
        ,{\hat \varrho}
        ]
        -
        z
        {
        D_{pp}  
        \over
         \hbar^2
        }
        {
        \lambda_M^2
        \over
                 4
        }
        \sum_{i=1}^3
        \frac i\hbar
        \left[  
        \left \{  
         {\hat {{\sf x}}}_i , 
        {\hat {\mbox{\sf p}}}_i
        \right \}            
        ,{\hat \varrho}
        \right]
        \nonumber
        \\
        &+&
        z{
        D_{pp}
        \over
         \hbar^2
        }
        \lambda_M^2
        \sum_{i=1}^3
        \left[  
        {
        {\hat {\mbox{\sf a}}}_i
        {\hat \varrho}
        {\hat {\mbox{\sf a}}}_i^{\scriptscriptstyle\dagger}
        - {\scriptstyle {1\over 2}}
        \{
        {\hat {\mbox{\sf a}}}_i^{\scriptscriptstyle\dagger}
        {\hat {\mbox{\sf a}}}_i
        ,{\hat \varrho}
        \}
        }
        \right]      ,
        \end{eqnarray}
where the single generator structure is due to the fact that the
coefficients in (\ref{coeff}) are actually linked by 
$
{D_{pp}}{D_{xx}} = 
{\hbar^2 {\gamma}^2 / 4}
$.
It is now also possible to consider the extension to quantum
statistics, thus obtaining a master equation describing the motion of
a Brownian particle in a Bose or Fermi gas. To do this one has to
calculate the dynamic structure factor 
of a free Bose or Fermi gas, taking then the
Brownian limit in which the ratio between the masses is much smaller
than one and the limit of small momentum transfer. The result one
obtains, when expressed in terms of the fugacity $z$ of the gas is in fact
particularly simple
        \begin{eqnarray}
        \label{1+z}
        {  
        d {\hat \varrho}  
        \over  
                dt  
        }  
        &=&
        -
        {i\over\hbar}
        [
        {{\hat {\mbox{\sf H}}}_0}
        ,{\hat \varrho}
        ]
        -
        {
        z
        \over
         1\mp z
        }
        \sum_{i=1}^3
        \left \{
        {
        D_{pp}
        \over
         \hbar^2
        }
        \left[  
        {\hat {{\sf x}}}_i,
        \left[  
        {\hat {{\sf x}}}_i,{\hat \varrho}
        \right]  
        \right]
        \right.
        \nonumber 
        \\
        &&
        \hphantom{\quad\quad}
        +
        \left.
        {
        D_{xx}
        \over
         \hbar^2
        }
        \left[  
        {\hat {\mbox{\sf p}}}_i,
        \left[  
        {\hat {\mbox{\sf p}}}_i,{\hat \varrho}
        \right]  
        \right]  
        +{i\over\hbar}
        \gamma
        \left[  
        {\hat {{\sf x}}}_i ,
        \left \{  
        {\hat {\mbox{\sf p}}}_i,{\hat \varrho}
        \right \}  
        \right]
        \right \}
        , 
        \end{eqnarray}
the operator structure is preserved, but the dependence on the
fugacity is through $z/ (1\mp z)$ rather than through $z$, thus
recovering for small fugacity the Boltzmann case (\ref{z}).
From  (\ref{1+z}) one also obtains a definite expression for the ratio 
between the  friction coefficient for the Maxwell Boltzmann and Bose
or Fermi statistics $$      
        {
        \gamma_{\rm \scriptscriptstyle MB}
        \over
        \gamma_{\rm \scriptscriptstyle B/F} 
        }=1\mp z,$$
showing that friction is enhanced or suppressed according to
statistics.

The author would like to thank Prof. L. Lanz for
useful discussions.
This work was  supported by MURST under Cofinanziamento and
Progetto Giovani.
%%%%%%%%%%%%%%%%%%%%%%%%%%%%%%%%%%%%%%%%%%%%%%%%%%%%%%%%%%%%%
  
\end{document}